\documentstyle[aaspp4,flushrt]{article}

\newcommand{\vect}[1]{\mbox{\boldmath $#1$}}   
\newcommand{\rr}{\mbox{\bf r}}
\newcommand{\R}{\mbox{\bf R}}
\newcommand{\n}{\mbox{\bf n}}
\newcommand{\ex}{\mbox{${\bf \hat{x}}$}}
\newcommand{\ey}{\mbox{${\bf \hat{y}}$}}
\newcommand{\vv}{\mbox{\bf v}}
\newcommand{\F}{\mbox{\bf F}}
\newcommand{\m}{\mbox{$\mu$}}
\newcommand{\md}{\mbox{$\acute{\mu}$}}
\newcommand{\ngc}{NGC~6712}

\begin{document}

\title{Tidal Shocking by Extended Mass Distributions}

\author{Oleg Y. Gnedin}
\affil{Princeton University Observatory, Princeton, NJ~08544; \\
       ognedin@astro.princeton.edu}

\author{Lars Hernquist$^1$}
\affil{University of California, Santa Cruz, CA~95064; \\
       lars@ucolick.org}

\author{Jeremiah P. Ostriker}
\affil{Princeton University Observatory, Princeton, NJ~08544; \\
       jpo@astro.princeton.edu}

\footnotetext[1]
{Presidential Faculty Fellow}

\begin{abstract}

We derive expressions for the tidal field exerted by a spherically symmetric
galaxy having an extended mass distribution, and use our analysis to
calculate tidal heating of stars in a globular cluster or a satellite
galaxy.  We integrate the tidal force over accurate orbits in various models
for the primary by using the impulse approximation, coupled with adiabatic
corrections that allow for time variation of the perturbation.  We verify
the results with direct N-body simulations, and also compare them with the
conventional straight path approximation.  Heating on highly eccentric
orbits dominates, as adiabatic corrections strictly prohibit energy changes
on low eccentricity orbits.  The results are illustrated for globular
cluster NGC~6712.  For the orbital eccentricities higher than 0.9, the
future lifetime of NGC~6712 is less than $10^{10}$ yr.  Our analysis can be
used to study tidal interactions of star clusters and galaxies in a
semi-analytical manner.

\end{abstract}

\keywords{globular clusters: general -- galaxies: interactions}

\section{Introduction}

Tidal forces are very important in stellar systems.  They determine the
dynamics of interactions between galaxies in clusters, between a dwarf
satellite and a primary galaxy, and between a star cluster and its host
galaxy.  The most dramatic tidal perturbations occur in fast and close
encounters.  When the duration of the encounter is shorter than the
characteristic dynamical time of the satellite, such an interaction is
referred to as a {\it tidal shock} (\cite{S:87}).

Many previous semi-analytic and numerical studies of tidal shocks considered
the primary galaxy as a point-mass (\cite{S:58}; \cite{R:75}; \cite{K:76}).
Such an approximation is violated when a satellite is well within the limits
of the galaxy or a globular cluster passes near the galactic nucleus.  More
recent work included the extended structure of the perturber (King and de
Vaucouleurs models in \cite{AW} 1985, 1986; isothermal sphere potential in
\cite{OL:92}; \cite{OLA:92}, 1995; and synthetic models in \cite{JSH:95};
\cite{Kr:97}), but most results were obtained through numerical simulations.
\cite{W:97} and \cite{MW} (1997a-c) used the linear perturbation theory and
semi-analytical Fokker-Planck calculations in their study of globular
clusters and elliptical galaxies.  The latter work is complex and cannot be
easily used for quick estimates.  In this paper, we obtain expressions for
the tidal field of an arbitrary spherically-symmetric system and apply them
to estimate the amount of heating of the satellite stars produced in the
interaction.  The resulting expressions are simple and ready-to-use.

The paper is organized so that we move from the simplest and least accurate
treatment to the more complex and more accurate one.  We present analytical
calculations of the tidal field and energy changes for straight-path orbits
in \S\ref{sec:straight} and for eccentric orbits in \S\ref{sec:eccentric}.
First, we use the impulse approximation and then relax it by allowing for
the motion of stars during the perturbation.  Some tedious integrals are
relegated to the Appendix.  Then, in \S\ref{sec:ngc}, we apply our theory to
calculate the heating of globular cluster \ngc\ in the Galaxy.  We discuss
and compare our method with previous work in \S\ref{sec:discussion}.  In
\S\ref{sec:summary}, we summarize our results for semi-analytic studies of
tidal interactions.


\section{Tidal perturbation in straight-path encounters}
\label{sec:straight}

Consider an inertial reference frame with the origin at the Galactic center.
Let $\R_0(t)$ be the trajectory of the center of the satellite, and $r$ the
distance of a member star from the satellite center.  Then the radius-vector
of a star is $\R = \R_0(t) + \rr$.  The equations of motion are
\begin{equation}
\ddot{\R} = -{\partial\Phi_c \over\partial\R}-{\partial\Phi_G \over\partial\R},
\end{equation}
or
\begin{equation}
\ddot{\R}_0 + \ddot{\rr} = - \left({\partial\Phi_G \over\partial\R_0}
   \right)_{R_0} - \left({\partial\Phi_c \over\partial\rr}\right)_R
   - \left({\partial^2 \Phi_G \over\partial\R_0 \partial\R_0}\right)_{R_0}
   \cdot \rr - ...
\end{equation}
where $\Phi_c$ and $\Phi_G$ are the potentials of the satellite and the
galaxy, respectively.  We will drop the subscripts henceforth.  The
trajectory of the center of mass of the satellite is determined by the
leading term, $\ddot{\R}_0 = - \partial\Phi_G / \partial\R_0$, so, in the
tidal approximation, we have
\begin{equation}
\ddot{\rr} = - {\partial \Phi_c \over \partial \rr}
   -{\partial^2 \Phi_G \over\partial\R_0\partial\R_0} \cdot \rr.
\end{equation}
The potential of a spherically-symmetric system is
\begin{equation}
\Phi_G = -{GM(R) \over R} - 4\pi G \int_R^\infty \, \rho(r') r' dr',
\end{equation}
which produces a tidal force per unit mass
\begin{equation}
\F_{\rm tid} \equiv -\left({\partial^2 \Phi_G \over\partial\R_0\partial\R_0}
  \right) \cdot \rr = {GM_0 \over R_0^3} \, \left[ (3\m - \md) (\n \cdot \rr) \n
  - \m \rr \right],
\label{eq:Ftid}
\end{equation}
where $M_0$ is the total mass of the galaxy, and $\m(R)$ is the normalized
mass profile:
\begin{equation}
\m(R) \equiv {M(R) \over M_0},
\end{equation}
and $\md$ is defined by
\begin{equation}
\md(R) \equiv {d\m(R) \over d\ln{R}}.
\end{equation}
The direction to the center of the satellite is denoted by $\n \equiv
\R_0/R_0$.

The treatment so far in this section is general.  We will now temporarily
restrict ourselves to the impulse approximation and straight line orbits for
simplicity in the computation.  In the impulse approximation, we neglect
internal motions of the satellite stars during the encounter. The
corresponding change in the star's velocity is the integral
\begin{equation}
\Delta \vv = \int \F_{tid} \, dt
\label{eq:dv}
\end{equation}
over the duration of the interaction. This time-varying perturbation leads
to an increase of the random motion (or heating) of stars in the
satellite. When the phase space is well mixed, the averaged energy change of
stars with the initial energy $E$ is quadratic in perturbation:
\begin{equation}
\langle \Delta E \rangle_E = \langle {1\over 2} \, (\Delta v)^2 \rangle.
\end{equation}
We have also neglected the self-consistent reaction of the satellite
potential as it settles into a new post-shock equilibrium.  Effects of the
potential readjustment are studied elsewhere (\cite{GO} 1998).

To calculate the velocity kick, $\Delta\vv$, we need to specify satellite's
orbit.  Since the tidal force is short-range, only the part of the orbit
within a few distances of the closest approach (perigalacticon $R_p$) is
important.  Therefore, the orbit is conventionally assumed to be a straight
path, $\R_0(t) = (R_p, V_p t, 0)$, where $V_p$ is the orbital velocity at
perigalacticon.  In \S\ref{sec:eccentric}, we consider eccentric orbits of
the satellite.

Integrating equation (\ref{eq:dv}) from $t = -\infty$ to $\infty$, we find
\begin{equation}
\Delta \vv = {2GM_0 \over R_p^2 V_p} \, \left\{ (3J_0-J_1-I_0)x, \,
             (2I_0-I_1-3J_0+J_1)y, \, -I_0 z \right\},
\end{equation}
where
\begin{mathletters}
\begin{eqnarray}
I_0(R_p) & \equiv & \int_1^\infty \, \m(R_p\zeta) \, {d\zeta \over \zeta^2 \,
           (\zeta^2-1)^{1/2}}, \\
I_1(R_p) & \equiv & \int_1^\infty \, \md(R_p\zeta) \, {d\zeta \over 
           \zeta^2 \, (\zeta^2-1)^{1/2}}, \\
J_0(R_p) & \equiv & \int_1^\infty \, \m(R_p\zeta) \, {d\zeta \over \zeta^4 \,
           (\zeta^2-1)^{1/2}}, \\
J_1(R_p) & \equiv & \int_1^\infty \, \md(R_p\zeta) \, {d\zeta \over 
           \zeta^4 \, (\zeta^2-1)^{1/2}}.
\end{eqnarray}
\label{eq:int}
\end{mathletters}
Notice also that
\begin{equation}
I_1 = R_p {dI_0 \over dR_p}, \hspace{1cm} J_1 = R_p {dJ_0 \over dR_p}.
\label{eq:deriv}
\end{equation}
Averaging over an ensemble of stars in a spherically-symmetric satellite, we
have $\langle x^2 \rangle = \langle y^2 \rangle = \langle z^2 \rangle =
r^2/3$.  The heating term per unit mass is therefore
\begin{eqnarray}
\langle \Delta E \rangle & = & \left( {2\, G \, M_0 \over R_p^2 \, V_p}
   \right)^2 \, {r^2 \over 3} \, \chi_{st}(R_p),
\label{eq:de} \\
\chi_{st} & \equiv & {1 \over 2} \, \left[ (3J_0-J_1-I_0)^2 +
                     (2I_0-I_1-3J_0+J_1)^2 + I_0^2 \right].
\label{eq:masscorr}
\end{eqnarray}
These results generalize the usual expressions for a point-mass
(e.g. \cite{BT} 1987) to the case of an arbitrary spherical density profile.
In particular, the function $\chi_{st}(R_p)$ is the correction due to the
extended mass distribution of the primary galaxy.  If all galactic mass was
concentrated at its center, then $\m=1$ and $\md=0$ everywhere, and we have
\begin{equation}
I_0 = \int_1^\infty {d\zeta \over \zeta^2 \, (\zeta^2-1)^{1/2}} = 1,
\hspace{1cm}
J_0 = \int_1^\infty {d\zeta \over \zeta^4 \, (\zeta^2-1)^{1/2}} = {2 \over 3},
\end{equation}
and $I_1 = J_1 = 0$. The factor $\chi_{st}$ becomes unity
in this case, as expected.  The subscript ``st'' reminds us that
this factor is valid only for the straight path approximation.

The second order energy diffusion term to the lowest order in perturbation
is
\begin{equation}
\langle \Delta E^2 \rangle_E = \langle (\vv \cdot \Delta\vv)^2 \rangle
  = \left( {2\, G \, M_0 \over R_p^2 \, V_p} \right)^2 \,
    {2\, r^2 \, v^2 \, (1+\chi_{r,v}) \over 9},
\end{equation}
where $\chi_{r,v}$ is the position-velocity correlation function, which
takes values from -0.25 to -0.57 (see \cite{GO} 1998).  This can be simply
rewritten as
\begin{equation}
\langle \Delta E^2 \rangle = {2\, v^2 \, (1+\chi_{r,v}) \over 3} \;
  \langle \Delta E \rangle.
\end{equation}
Higher order terms, $\langle \Delta E^n \rangle$, are negligible for tidal
shocks (\cite{S:87}).  For example, both $n=3$ and $n=4$ terms are quadratic
in the perturbation energy, $\Delta E/E \ll 1$.  Therefore, only the first
two terms are important in the Fokker-Planck equation (see
\S\ref{sec:discussion}).

\subsection{The Hernquist model}

\begin{figure} \plotone{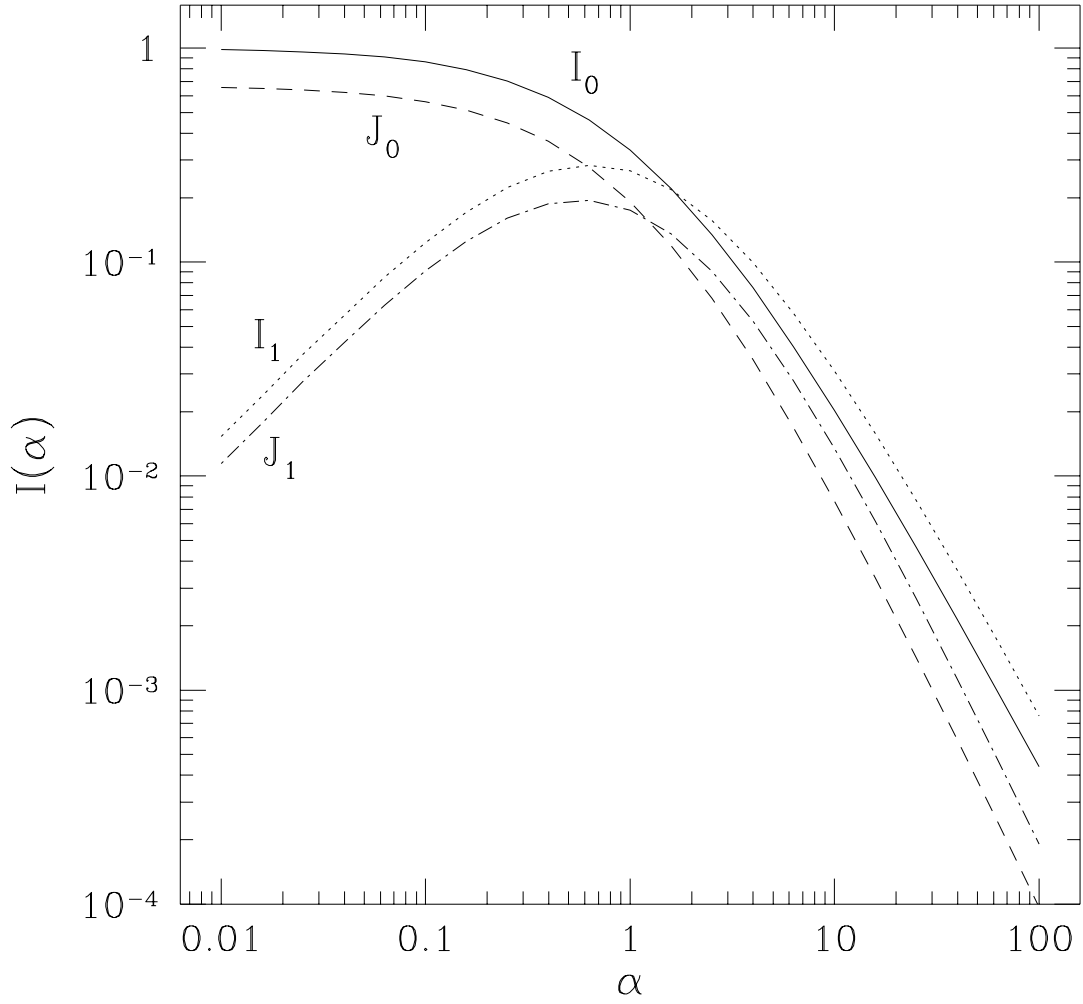}
\caption{Integrals $I_0$, $J_0$, $I_1$, and $J_1$ (eq.
  [\protect\ref{eq:Hint}]) for the Hernquist model.
  \label{fig:Hint}}
\end{figure}

For illustration, we consider several examples of mass distributions, the
Hernquist model (\cite{H:90}) and power-law density profiles.  The Hernquist
model closely resembles the $R^{1/4}$ law for spherical galaxies and gives
simple analytical expressions for the potential-density pair:
\begin{equation}
\Phi = - {G M_0 \over R+a}, \hspace{1cm}
\rho = {M_0 a \over 2\pi R (R+a)^3},
\end{equation}
where $M_0$ is the total mass of the galaxy, and $a$ is the scale length.
The normalized mass profile is
\begin{equation}
\m(R) = {R^2 \over (R+a)^2}, \hspace{1cm}
\md(R) = {2a R^2 \over (R+a)^3}.
\label{eq:mbH}
\end{equation}
The calculation of the integrals (\ref{eq:int}) is given in Appendix A.  The
results (eq. [\ref{eq:Hint}]) are plotted in Figure \ref{fig:Hint} versus
the dimensionless parameter $\alpha \equiv a/R_p$.

For large impact parameters ($\alpha \ll 1$), both $I_0$ and $J_0$ approach
constant values ($I_0=1$, $J_0=2/3$), while $I_1$ and $J_1$ decrease
linearly with $\alpha$.  As a result, the satellite is compressed in the
vertical direction, stretched in the direction of the perigalacticon by the
same amount, and unaffected in the perpendicular direction.  All this is
expected in the point-mass approximation.  At very small impact parameters
($\alpha \gg 1$), when the satellite penetrates within a scale length of the
center, all correction factors decline approximately as $\alpha^{-2}$.  This
asymptotic behavior is due to the mass of the model increasing quadratically
with the radius at small $r$.

\begin{figure} \plotone{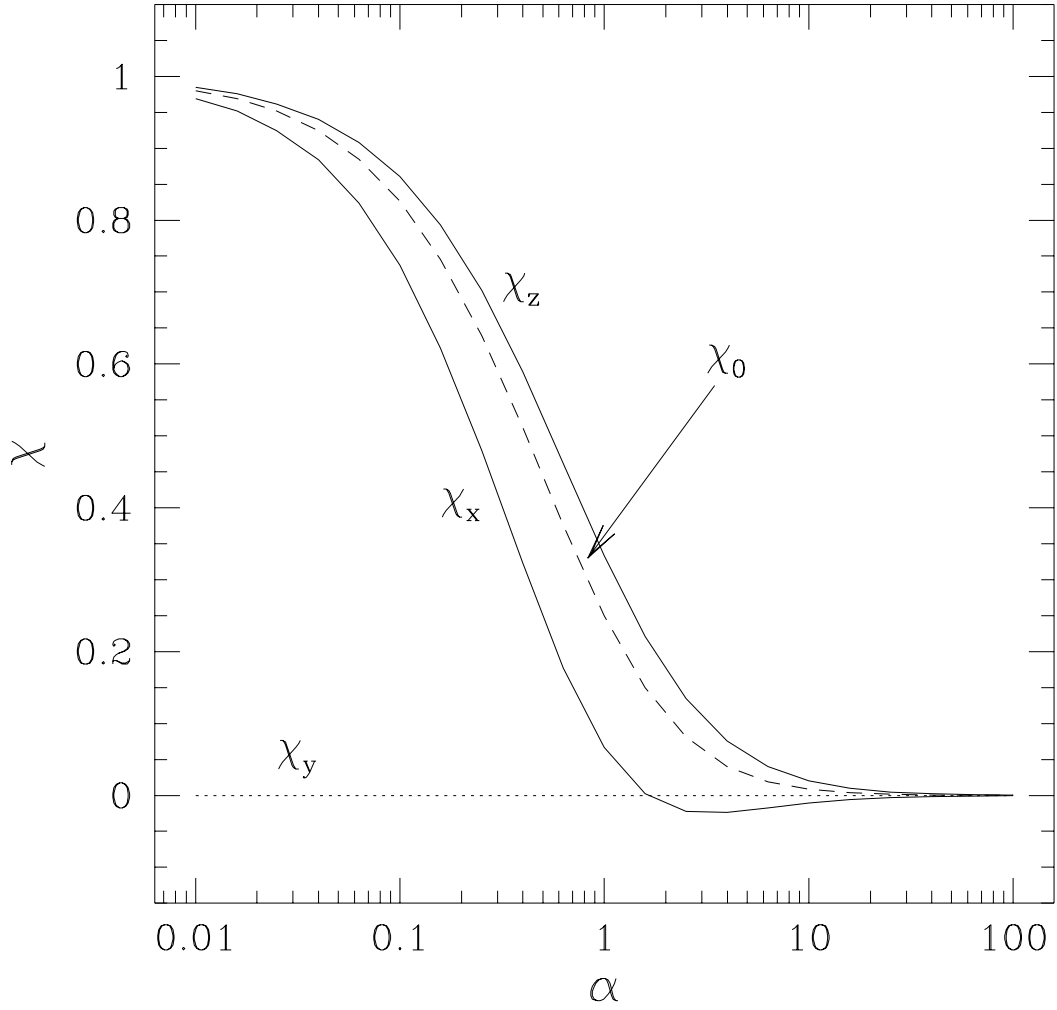}
\caption{Corrections to the velocity change relative to the point-mass
  approximation for the Hernquist model (cf eq. [\protect\ref{eq:chivel}]).
  The value of $\chi_y$ is always close to zero, as it would be in a
  point-mass potential.  For comparison, dashes show the fraction of the
  galactic mass inside the perigalacticon, $\chi_0 = \m(R_p)$.
  \label{fig:chivel}}
\end{figure}

In Figure \ref{fig:chivel}, we plot the absolute value of the components of
the velocity change for the Hernquist model:
\begin{mathletters}
\begin{eqnarray}
\chi_x & \equiv & 3 J_0 - J_1 - I_0 \\
\chi_y & \equiv & 2 I_0 - I_1 - 3 J_0 +J_1 \\
\chi_z & \equiv & I_0
\end{eqnarray}\label{eq:chivel}
\end{mathletters}
Note that $\chi_x$ and $\chi_z$ are the velocity changes in the $x$ and $z$
directions relative to the point-mass approximation.  The velocity change in
the $y$ direction is negligible for any $\alpha$; its typical value is
$\chi_y \sim 10^{-4}$.  In the point-mass approximation, $\Delta v_y \equiv
0$.  Also, we plot for comparison a naive approximation, which assumes the
galactic mass inside the perigalacticon $R_p$ to be a point-mass. The
correction factor in this case is $\chi_{x0} = \chi_{z0} = \chi_0 \equiv
\m(R_p)$.  It underestimates the vertical compression, but overestimates the
effect in the radial direction.

\subsection{The power-law models}

Next, we consider a general power-law distribution truncated at a limiting
radius $R_{max}$,
\begin{equation}
\rho(R) = \left\{ \begin{array}{ll}
                   k\, R^{-n}, \;\;\; & R \leq R_{max} \;\; (n<3) \\
                   0,                 & R > R_{max}
                  \end{array} \right.
\end{equation}
Here $k$ is a constant which determines the total mass of the model, $M_0 =
{4\pi k \over 3-n} R_{max}^{3-n}$.  We assume that the perturbed satellite
is well within the extent of the galaxy, so that truncation of the model
does not affect our results.  The normalized mass profile is
\begin{equation}
\m(R) = \left( {R \over R_{max}} \right)^{3-n}, \;\;\; \md(R) = (3-n) \, \m(R).
\label{eq:powermass}
\end{equation}
The evaluation of the tidal integrals (eqs. [\ref{eq:int}]) is
straightforward:
\begin{mathletters}
\begin{equation}
I_0 = \left( {R_p \over R_{max}} \right)^{3-n} \, {\sqrt{\pi} \over 2} \,
   {\Gamma{\left( {n-1 \over 2}\right)} \over\Gamma{\left( {n \over 2}\right)}},
   \;\;\;\;  I_1 = (3-n) \, I_0.
\end{equation}
\begin{equation}
J_0 = \left( {R_p \over R_{max}} \right)^{3-n} \, {\sqrt{\pi} \over 2} \,
   {\Gamma{\left( {n+1\over 2}\right)} \over\Gamma{\left( {n+2\over 2}\right)}},
   \;\;\;\;  J_1 = (3-n) \, J_0.
\end{equation}
\end{mathletters}
Note that the integral $I_0$ converges only for $n>1$. Using the
identity $\Gamma(x+1) = x\Gamma(x)$, we find $n\, J_0 = (n-1)\, I_0$.

The change of the stellar velocity is
\begin{equation}
\Delta \vv = {GM_0 \over R_p^2 V_p} \, \left( {R_p \over R_{max}} \right)^{3-n}
  \, {\sqrt{\pi} \, \Gamma{\left( {n-1 \over 2}\right)} \over
  \Gamma{\left( {n \over 2}\right)}} \; \left\{ (n-2)x, \, 0, \, -z \right\},
\end{equation}
where the $y$-component vanishes because of the symmetry of the trajectory.
The correction to the energy change is
\begin{equation}
\chi_{st} = \left( {R_p \over R_{max}} \right)^{6-2n} \, \left[
   {\Gamma{\left( {n-1\over 2}\right)} \over\Gamma{\left( {n\over 2}\right)}}
   \right]^2 \, {\pi \, (n^2-4n+5) \over 8}.
\end{equation}
We remind the reader that this result holds for the power-law indices $1 < n
< 3$.

Notice the change in sign of $\Delta v_x$ as the index $n$ passes
through the value $n=2$.  For the special case of an isothermal sphere
($n=2$), the $x$-component of the velocity change vanishes identically, and
the only resulting change is compression in the vertical direction:
\begin{equation}
\Delta \vv = - {G M_0 \over R_p^2 V_p} \, {\pi R_p \over R_{max}} \,
  \vect{z}, \hspace{1cm}
\chi_{st}(R_p) = {\pi^2 \over 8} \, {R_p^2 \over R_{max}^2}.
\end{equation}


\section{Tidal perturbation on eccentric orbits}
\label{sec:eccentric}

The straight-path orbit is a convenient approximation, but in real galaxies,
satellites move on eccentric orbits.  Effects of tidal shocks are most
prominent in the vicinity of the perigalacticon.  The results obtained in
the previous section should nevertheless give us insight into the actual
heating of satellites on eccentric orbits.  In this section, we determine
the true orbits in our examples of galactic models and investigate how tidal
heating depends on parameters of the orbit.

In a spherically-symmetric potential, the orbit of a satellite is confined
to a plane and can be characterized by two independent parameters.  Let us
choose the perigalactic distance, $R_p$, and eccentricity, $e$, as such
parameters.  They in turn determine the energy and angular momentum of the
orbit.  We employ polar coordinates, in which the position angle $\theta(t)$
can be used as a time variable.  This angle is defined such that $\theta=0$
at $R=R_p$.  In the coordinate frame where the orbit is in the XY plane, the
position vector of the satellite is $\n(t) = \ex \cos{\theta(t)} + \ey
\sin{\theta(t)}$.

Using the impulse approximation, we integrate the tidal force along the
orbit to obtain the expected change in the stellar velocity:
\begin{equation}
\Delta \vv = \int_{-T/2}^{T/2} \F_{\rm tid} \, dt,
\end{equation}
where $T$ is the orbital period.

Changing the variable of integration from time, $t$, to angle, $\theta$,
using the identity $dt = (R^2/J) \, d\theta$, where $J$ is the orbital
angular momentum, we obtain
\begin{equation}
\Delta \vv = \left({G M_0 \over a^3}\right)^{1/2} \,
  \int_{-\theta_m}^{\theta_m} \, d\theta \,
  {a \over j \, R} \left[ \left( 3\m-\md \right) (\n \cdot \rr)\n
  - \m\rr \right],
\end{equation}
where $a$ is the parameter of the galactic model ($R_{max}$ in case of the
power-law models), and $j$ is the dimensionless angular momentum, defined by
$J \equiv (G M_0\, a)^{1/2} j$.  The integration over the position angle
extends to the maximum angle $\theta_m$ corresponding to the apogalactic
point of the orbit.  Thus, the tidal force is integrated starting from the
apogalacticon at $\theta = -\theta_m$, approaching the perigalacticon at
$\theta=0$, and going to the next apogalacticon at $\theta = \theta_m$.  The
maximum value of the position angle ranges from $\theta_m = \pi$ in the
Keplerian potential to $\theta_m = \pi/2$ in the harmonic oscillator
potential.  A very accurate procedure for calculating the orbits is
described in Appendix B.

Rewriting the velocity change into components, we have
\begin{equation}
\Delta \vv = \left({G M_0 \over a^3}\right)^{1/2} \, {1 \over j(\alpha,e)} \,
   \left\{ (B_1-B_3) x, \, (B_2-B_3) y, \, -B_3 z \right\},
\end{equation}
where
\begin{mathletters}
\begin{eqnarray}
B_1 & = & \int_{-\theta_m}^{\theta_m} \, {3\m(R)-\md(R) \over R/a} \,
          \cos^2{\theta} \, d\theta \\
B_2 & = & \int_{-\theta_m}^{\theta_m} \, {3\m(R)-\md(R) \over R/a} \,
          \sin^2{\theta} \, d\theta \\
B_3 & = & \int_{-\theta_m}^{\theta_m} \, {\m(R) \over R/a} \, d\theta.
\end{eqnarray}
\label{eq:B}
\end{mathletters}
The cross-terms ($\propto xy$) vanish because of the symmetry of the orbit,
$R(\theta) = R(-\theta)$.  The energy changes of stars with the initial
energy $E$ are
\begin{eqnarray}
\langle\Delta E\rangle_E & = & {G M_0 \over a^3} \, r^2 \,
   {(B_1-B_3)^2 + (B_2-B_3)^2 + B_3^2 \over 6 j^2(\alpha,e)} \, A_1(x_t).
   \label{eq:dEecc} \\
\langle\Delta E^2\rangle_E & = & {G M_0 \over a^3} \, {2\, r^2 \, v^2 \,
   (1+\chi_{r,v}) \over 3} \,
   {(B_1-B_3)^2 + (B_2-B_3)^2 + B_3^2 \over 6 j^2(\alpha,e)} \, A_2(x_t).
   \label{eq:dE2ecc}
\end{eqnarray}

We have now added adiabatic corrections, $A_1(x_t)$ and $A_2(x_t)$, which
allow for the motion of stars during the perturbation.  Adiabatic
corrections reduce the amount of heating as the result of the conservation
of stellar adiabatic invariants.  \cite{GO} (1998) provide a detailed
discussion of the adiabatic corrections in the case of disk shocking.  The
main results can be summarized as follows.  The change of the stellar energy
inferred from the N-body modeling of tidal shocks is described to a good
accuracy by the product of the impulsive value and the adiabatic correction
of the form
\begin{equation}
A_i (x_t) = (1 + x_t^2)^{-\gamma_i},
\label{eq:adcorr}
\end{equation}
where $x_t \equiv \omega \tau$.  Here $\omega(r)$ is the orbital frequency
of stars in the satellite, and $\tau$ is the effective duration of the
shock.  The exponents $\gamma_i$ depend on the shock duration relative to
the half-mass dynamical time of the satellite, $t_{dyn}$, and vary from
($\gamma_1 = 2.5$, $\gamma_2 = 3$) for $\tau \lesssim t_{dyn}$ to ($\gamma_1
= \gamma_2 = 1.5$) for $\tau \gtrsim 4\, t_{dyn}$.  The adiabatic correction
becomes increasingly small in the satellite core, conserving stellar actions
and energy.

The simulations were done for the King model of the satellite, with the
concentration parameter $c = 0.84$.  We have checked that expression
(\ref{eq:adcorr}) remains valid for most parts of the more concentrated
model with $c = 1.5$.  However, our adiabatic corrections slightly
underestimate the energy change in the core.  It is possible that equation
(\ref{eq:adcorr}) breaks for very concentrated satellites and thus should be
applied with caution.  See also \cite{W:94} (1994a-c) for the direct
application of the perturbation theory formalism.

The adiabatic correction was calculated under the assumption that the
perturbing force varies with time as a Gaussian function, $F_{\rm tid}
\propto e^{-t^2/\tau^2}$.  This assumption does not strongly limit the
applicability of the results since numerical simulations show that
significant energy changes occur only close to the perigalacticon, where
almost any tidal force has similar behavior.  As an example, consider a
Keplerian orbit of eccentricity $e$.  Near the perigalacticon, the angle
$\theta \ll 1$ and the vertical component of the tidal force varies in
agreement with the Gaussian: $F_{zz} \approx F_{zz,0} (1 - c_1 \theta^2)
\approx F_{zz,0} (1 - c_2 t^2)$, where $c_1$ and $c_2$ are some constants.
See \S\ref{sec:ngcorbit} for more discussion of the temporal structure of
the tidal force.

The adiabatic correction depends not only on the stellar frequency,
$\omega$, but also on the orbital eccentricity of the satellite, through the
parameter $\tau$.  On low eccentricity orbits, the effective duration of the
``shock'' becomes so long that adiabatic corrections prohibit the changes of
the stellar energy.  On circular orbits, the effects of tidal shocks are
greatly reduced.  We will return to adiabatic corrections in
\S\ref{sec:ngc}, when we consider the evolution of globular cluster \ngc.

\subsection{The Hernquist model}

It is instructive to compare the results for the satellites on eccentric
orbits with those on the straight path.  For the Hernquist model, we perform
the integration of equations (\ref{eq:B}) numerically for a grid of
perigalactica and eccentricities.  We now rewrite equation (\ref{eq:dEecc})
for $\langle\Delta E\rangle$ similarly to equation (\ref{eq:de}), extracting
the factor $\chi_{ecc}$ which is now a function of the two parameters,
$\alpha \equiv a/R_p$ and $e$.  For clarity, the factor
$\chi_{ecc}(\alpha,e)$ does not include the adiabatic correction $A_1(x_t)$.

\begin{figure}
\plotone{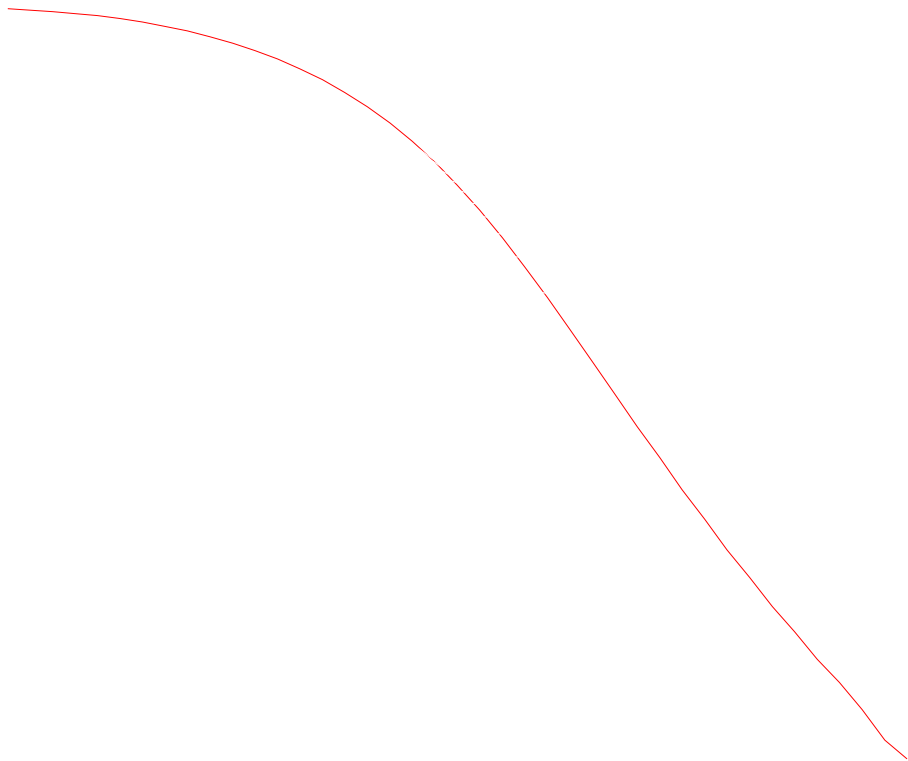}
\caption{The ratio of the correction factors, $\chi_{ecc}/\chi_{st}$, for
  the eccentric and straight orbits of the satellite.  The lines correspond to
  the orbits with eccentricities running from $e=0$ to 1 with a step of 0.1,
  in the Hernquist model for the galaxy.  Here $\alpha \equiv a/R_p$, where
  $a$ is the scale length of the model and $R_p$ is the perigalacticon of the
  orbit.  In the calculation of each curve, we assumed the same angular
  momentum of the straight orbit as for the corresponding eccentric orbit.
  Accordingly, the value of $\chi_{st}$ may vary for each line.  Adiabatic
  corrections are not included.  Had they been, the correction factors for $e
  \rightarrow 0$ would have been reduced dramatically.
  \label{fig:corr_h}}
\end{figure}

Figure \ref{fig:corr_h} shows the ratio $\chi_{ecc}/\chi_{st}$ for the
orbits with various eccentricities.  For highly eccentric orbits, $e
\lesssim 1$, the ratio varies slightly with $\alpha$, from
$\chi_{ecc}/\chi_{st} \approx 1.8$ for $\alpha \ll 1$ to
$\chi_{ecc}/\chi_{st} = 1$ for $\alpha \gg 1$.  The latter limit corresponds
to the most eccentric Keplerian orbit where both approaches must agree.
However, as the eccentricity decreases, the ratio $\chi_{ecc}/\chi_{st}$
varies with $\alpha$ much more strongly.  While at $\alpha \ll 1$, the ratio
is highest for $e=0$, the opposite is true for $\alpha \gg 1$.

\subsection{The power-law models}

The power-law models are scale-free and allow further simplification of
equations (\ref{eq:B}).  We have $B_i(\alpha,e) \equiv \alpha^{n-2} \,
C_i(e)$, $j^2(\alpha,e) \equiv \alpha^{n-4} \, f(e)$, and therefore
\begin{eqnarray}
\langle \Delta E \rangle & = & {G \, M_0 \over R_{max}^3} \, r^2 \,
    \alpha^n \, \chi_{ecc}(e) \, A_1(x_t),
\label{eq:dEpower} \\
\chi_{ecc} & \equiv & {(C_1-C_3)^2 + (C_2-C_3)^2 + C_3^2 \over 6 f(e)}.
\label{eq:chipower}
\end{eqnarray}
The remaining integrals depend only on the eccentricity $e$.  After some
algebra, we obtain
\begin{mathletters}
\begin{eqnarray}
C_1 & = & n \, \int_{-\theta_m}^{\theta_m} \, \left({R \over R_p}\right)^{2-n}
          \, \cos^2{\theta} \, d\theta, \\
C_2 & = & n \, \int_{-\theta_m}^{\theta_m} \, \left({R \over R_p}\right)^{2-n}
          \, \sin^2{\theta} \, d\theta, \\
C_3 & = & \int_{-\theta_m}^{\theta_m} \, \left({R \over R_p}\right)^{2-n} \, 
          d\theta.
\end{eqnarray}
\label{eq:C}
\end{mathletters}

Note that in the harmonic oscillator potential ($n=0$), the first two
integrals are identically zero: $C_1 = C_2 = 0$.  The third one can be done
using simple analytic orbits allowed in this potential
(eq. [\ref{eq:A.oscorb}]):
\begin{equation}
C_3 = \pi \, {1+e \over 1-e}, \;\;\;\;\;\;
\chi_{ecc} = {\pi^2 \over 2} \;\;\;\; (\mbox{for }n=0).
\end{equation}
Surprisingly enough, the change of the stellar energy depends neither on the
location in the galaxy, $\alpha$, nor on the orbital eccentricity, $e$.  The
satellite is equally compressed in all directions:
\begin{eqnarray}
\Delta \vv & = & \left({G M_0 \over R_{max}^3}\right)^{1/2} \, \pi \,
  \{ -x, -y, -z \}, \\
\langle \Delta E \rangle & = & {G M_0 \over R_{max}^3} \, {\pi^2 \over 2} \,
  r^2 \, A_1(x_t) \;\;\;\; (\mbox{for }n=0).
\end{eqnarray}

In the isothermal sphere potential ($n=2$), the integrals are also done
straightforwardly:
\begin{mathletters}
\begin{eqnarray}
C_1 & = & 2 \theta_m + \sin{2 \theta_m}, \\
C_2 & = & 2 \theta_m - \sin{2 \theta_m}, \\
C_3 & = & 2 \theta_m, \;\;\;\;\;\; (\mbox{for }n=2).
\end{eqnarray}
\end{mathletters}

\begin{figure}
\plotone{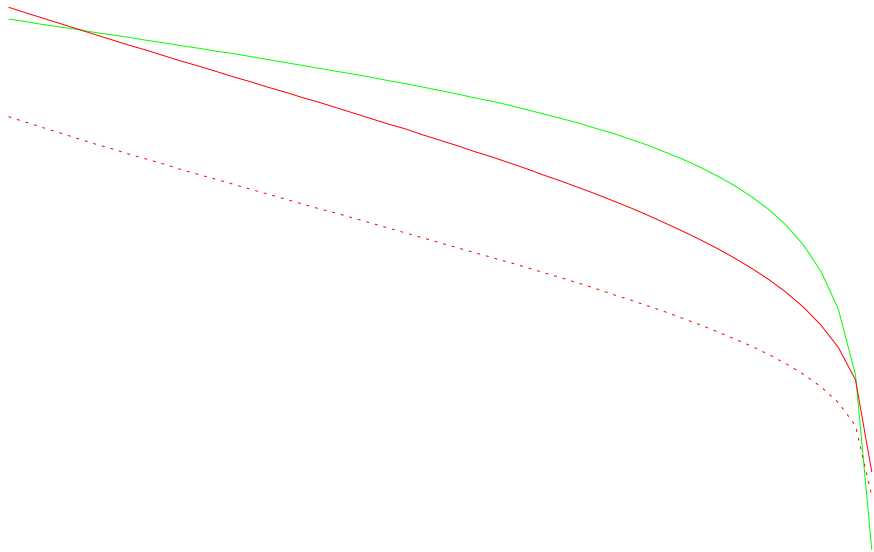}
\caption{The correction factor $\chi_{ecc}$ for the power-law models
  (eq. [\protect\ref{eq:chipower}]) versus the eccentricity of the orbit
  (solid lines).  For the isothermal sphere ($n=2$), we also show the
  straight-path approximation, $\chi_{st}$ (dots).
  \label{fig:power}}
\end{figure}

For the $n=1$ model, the integrals $C_i(e)$ are done numerically.  Figure
\ref{fig:power} shows the factors $\chi_{ecc}$ for all three models.  For
comparison, we also plot $\chi_{st}$ for the isothermal sphere model.
Heating on eccentric orbits is larger than that on the straight path, except
for $e=1$, where the two are equal.  However, the plot does not include
adiabatic corrections, which would have effectively reduced heating on
low-eccentricity orbits.

\section{A Case Study: \ngc}
\label{sec:ngc}

We now illustrate the theory developed in \S\ref{sec:eccentric} by
considering the evolution of a globular star cluster, \ngc.  This cluster is
presently at the distance of 3.8 kpc from the Galactic center, close enough
to experience tidal shocking by the Galactic bulge.  We use the observed
parameters to construct cluster's orbit in the bulge potential and to
integrate the expected energy change.  Then, we perform N-body simulations
to verify the result.  Finally, we consider various orbital eccentricities
allowed by the cluster orbital energy and illustrate the eccentricity effect
on the resulting heating.

\subsection{Constructing the orbit}
\label{sec:ngcorbit}

Full three-dimensional kinematic data for \ngc\ are not yet available.
Instead, we use the results of Monte-Carlo simulations by \cite{GO:97} for
the Ostriker-Caldwell (1983) model of the Galaxy with the isotropic velocity
distribution of the globular cluster system.  We draw the two unknown
components of the orbital velocity consistent with the chosen kinematic
model and integrate the orbits for $10^{10}$ yr.  For \ngc, the median
perigalactic distance is $R_p = 0.73$ kpc and the velocity at perigalacticon
is $V_p = 396$ km s$^{-1}$.  The eccentricity of this orbit is $e = 0.69$.

We approximate the Galactic bulge by a Hernquist model with the scale-length
$a = 0.6$ kpc.  The total mass of the model required to reproduce the
observed velocity $V_p$ is $M_0 = 3.3\times 10^{10}\, M_{\sun}$, close to
the observed mass of the bulge.

\begin{figure}
\plotone{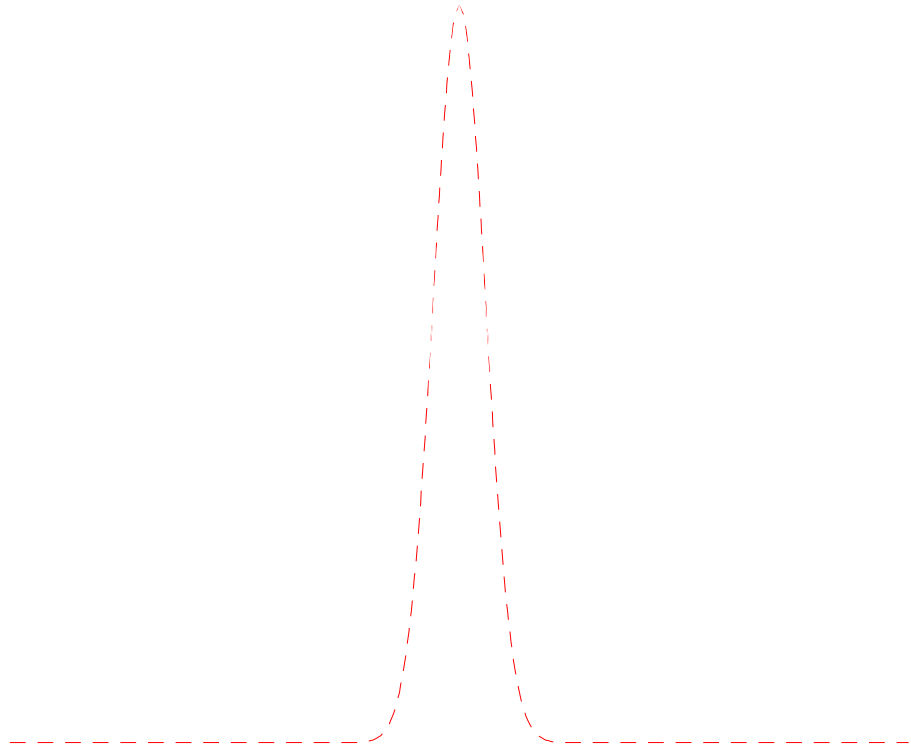}
\caption{Time variation of the vertical component of the tidal force
  exerted on globular cluster \ngc\ by the Galactic bulge.  Solid line
  is the actual force, whereas dashes show the Gaussian fit with
  the effective width $\tau = 5.3\, t_{dyn}$.  The force is in units
  of the N-body code.
  \label{fig:ngc_tidalforce}}
\end{figure}

Figure \ref{fig:ngc_tidalforce} shows the vertical component of the tidal
force exerted on the cluster along a single orbit in the Galaxy.  The force
peaks at the perigalacticon.  The dashed line shows that the Gaussian fit to
the actual force, $F_{fit} \propto e^{-t^2/\tau^2}$, is accurate for the
most important part of the orbit.  (A similar parameterization was employed
by \cite{JHW:98} in their numerical study of tidal heating.)  The width of
the Gaussian, or the effective duration of the shock, is $\tau \approx 5.3\,
t_{dyn}$, where $t_{dyn}$ is the half-mass dynamical time of the cluster.
This timescale is long enough that most stars in the cluster are protected
by adiabatic invariants, even for $R_p < 1$ kpc.  Using the calculated value
of $\tau$, we estimate adiabatic corrections to the energy change
(cf. eq. [\ref{eq:adcorr}]) and verify them with N-body simulations
described next.

\subsection{Self-Consistent N-body simulations}

We can now simulate the cluster as a system of $N$ point-mass particles and
run it along the chosen orbit.  At the end, we calculate the changes of the
stellar energies and compare them with our analytic theory.

We use a Self-Consistent Field code (\cite{HO:92}, \cite{GO} 1998) with $N =
10^6$ particles.  The code approximates the true gravitational potential and
density of the cluster with a finite series of basis functions.  In this
work, we take $n=10$ radial functions and $l=6$ spherical harmonics.

The initial cluster realization is a King model with the concentration
parameter $c = 0.84$.  It is sufficiently close to the observed
concentration of \ngc, $c \approx 0.9$.  The code employs units in which
$G=1$, $M_{cl}=1$, and $R_c=1$, where $M_{cl}$ and $R_c$ are the mass and
the initial core radius of the cluster.  The physical units are fixed by the
observed values, $M_{cl} = 2.5\times 10^5\, M_{\sun}$ and $R_c = 1.9$ pc.

In this simulation, the gravitational potential of the cluster is kept
fixed.  The self-consistent response of the system to tidal perturbations is
studied elsewhere (\cite{GO} 1998) and is not important for our discussion.
We run the cluster along the orbit, starting from the apogalacticon, where
the tidal force is minimal (cf. Figure \ref{fig:ngc_tidalforce}), going to
the perigalacticon, and then moving to the next apogalacticon.  The whole
simulation lasts for about 200 dynamical times, $t_{dyn}$, of the cluster,
with a short enough time step to ensure accurate integration of stellar
orbits; $\Delta t = 0.01\, t_{dyn}$.  We have run the simulation on the SGI
Origin 2000 supercomputer at the Princeton University Observatory.

\begin{figure}
\plotone{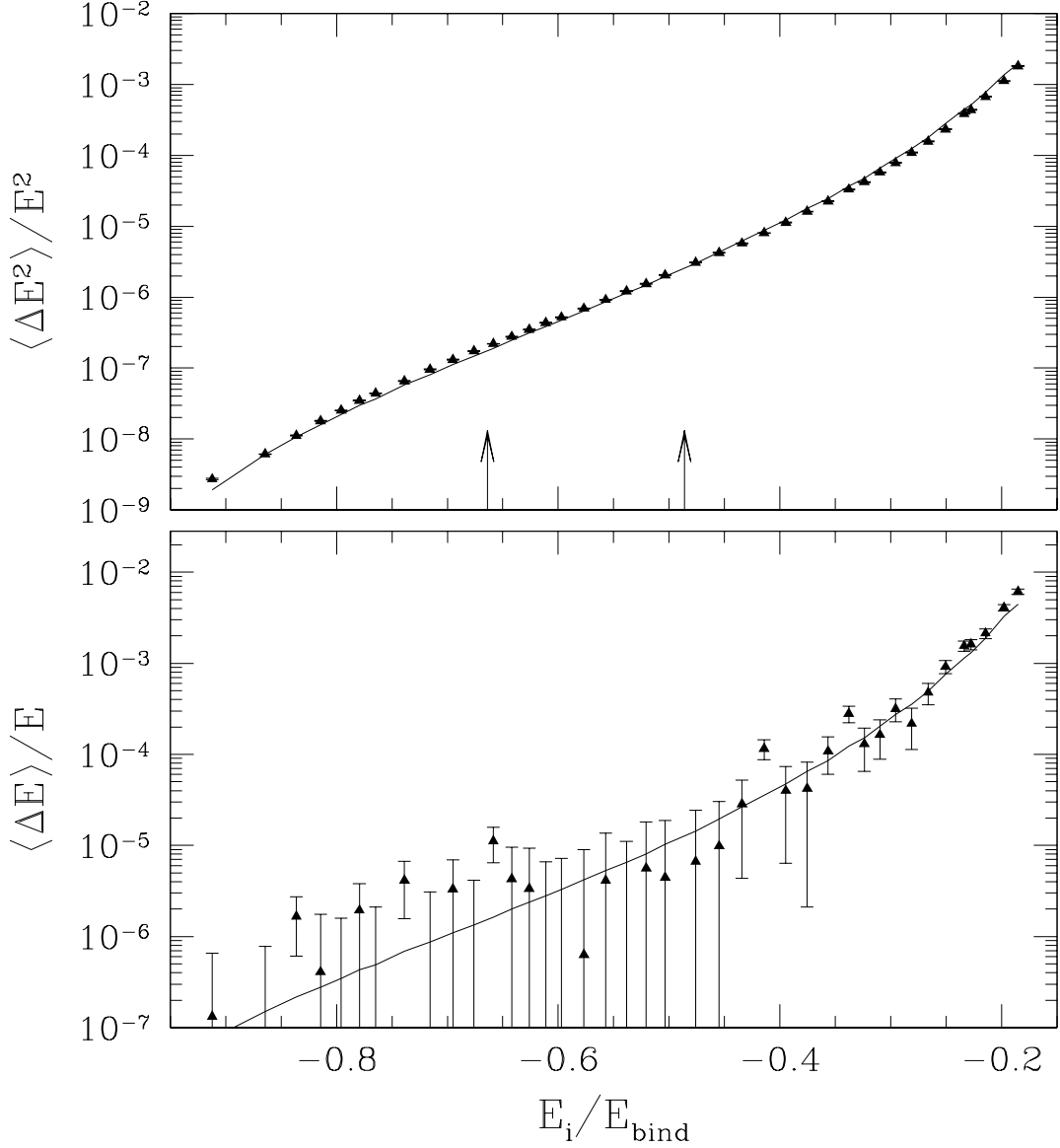}
\caption{The first and second order energy changes for globular cluster
  \ngc\ (triangles).  Both changes are plotted versus the initial energy of
  stars, $E_i$, and are normalized to the corresponding power of $E_i$.
  Error-bars indicate standard deviations in each energy bin.  Here
  $E_{bind}$ is the initial binding energy of the system, and arrows are at
  the core and the half-mass radii.  Solid lines are the analytic predictions
  (eq. [\protect\ref{eq:dEecc}]).
  \label{fig:ngc_bin}}
\end{figure}

Figure \ref{fig:ngc_bin} shows the first and second order energy changes at
the end of the simulation.  The particles are grouped in equal size bins,
ranked by their initial energy, and the mean values, $\langle\Delta
E\rangle$ and $\langle\Delta E^2\rangle$, and their standard errors are
calculated for each bin.  The data are compared with our analytic result
(eqs. [\ref{eq:dEecc},\ref{eq:dE2ecc}]), wherein adiabatic corrections
(eq. [\ref{eq:adcorr}] with $\gamma_1 = 1.4$, $\gamma_2 = 1.5$) are
appropriate for the long duration of the shock, $\tau$, estimated in
\S\ref{sec:ngcorbit}.  Had we used the simplest theory, presented in
\S\ref{sec:straight}, for straight line orbits without adabatic corrections,
we would have overestimated the total energy change significantly.  By
calculating the correct orbits and employing adiabatic corrections, we find
a very good agreement with the simulations.  This excellent agreement
encourages one to use the semi-analytical theory
(eqs. [\ref{eq:dEecc}-\ref{eq:adcorr}]) to study the effects of tidal shocks
on the evolution of star clusters and satellite galaxies.

\subsection{Effects of orbital eccentricity}

Finally, we demonstrate how the eccentricity of the orbit affects the
resulting heating.  We keep the orbital energy of \ngc\ fixed and consider
all allowed eccentricities.  Since these two parameters ($E_{orb}$ and $e$)
specify the orbit, the perigalactic distance $R_p$ varies accordingly with
$e$.  Increasing eccentricity decreases $R_p$ and vice versa.  Also, low
eccentricity orbits have longer shock durations, $\tau$, which leads to
significant adiabatic corrections.

\begin{figure}
\plotone{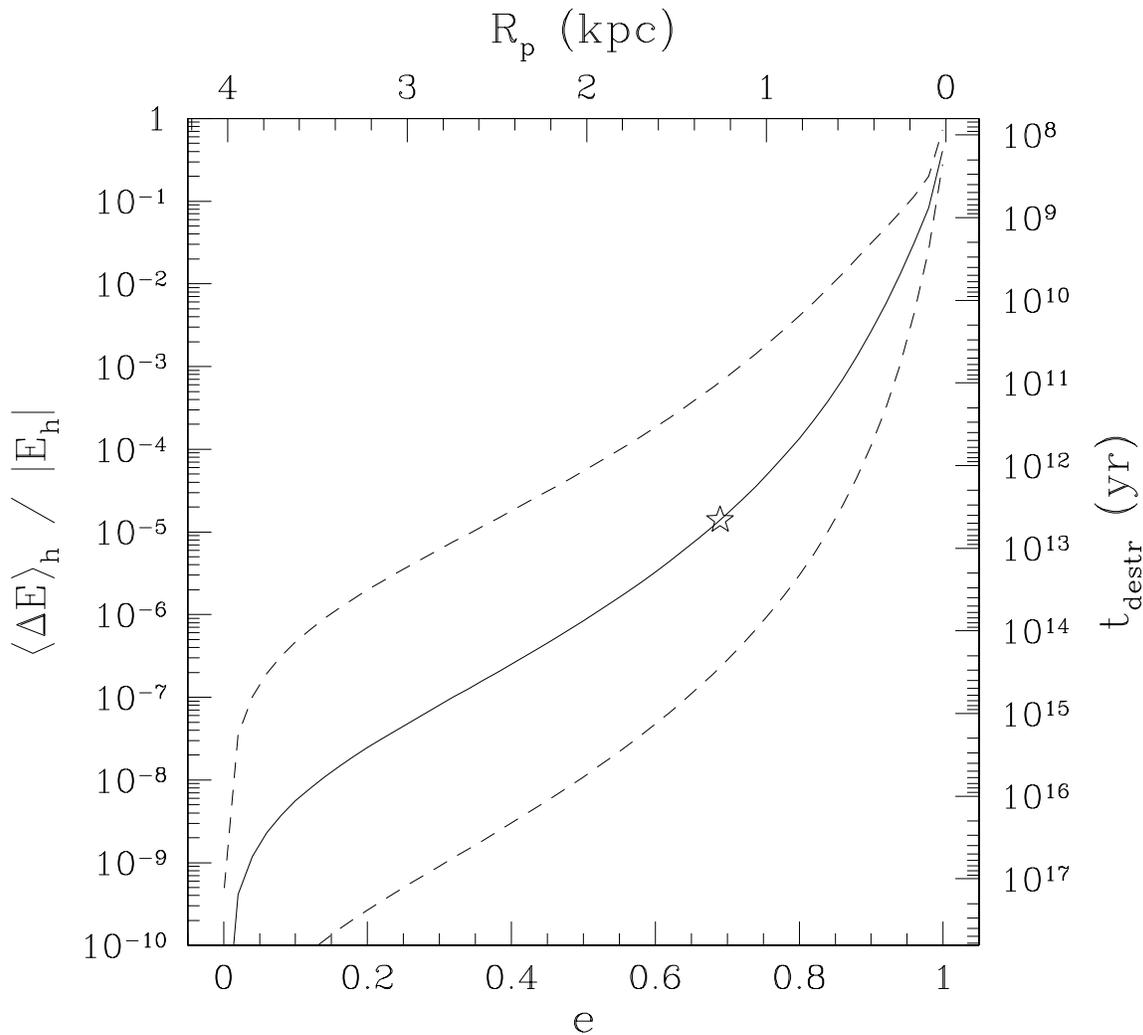}
\caption{Energy change of stars at the half-mass radius of \ngc\
  relative to the initial half-mass energy as a function of eccentricity
  of the orbit, $e$, and the corresponding perigalactic distance, $R_p$.
  The orbital energy is fixed at the observed value for
  the cluster (solid line) or at one-half and twice that value
  (dashed lines below and above the solid line, respectively).
  The star marks the eccentricity of \ngc\ in our kinematic model
  (see \S\protect\ref{sec:ngcorbit}).
  For illustration, we also show on the right-hand-side axis
  the approximate destruction time of the cluster due only to tidal
  shocks.  The destruction time is defined as
  $t_{destr} \equiv P_{orb} (E_h / \Delta E_h)$, where $P_{orb}$ is
  the cluster orbital period, which varies with eccentricity.
  In contrast to Figure \protect\ref{fig:corr_h}, which did not
  allow for adiabatic corrections, here tidal shocking
  vanishes as $e \rightarrow 0$.
  \label{fig:ngc_ecc}}
\end{figure}

Figure \ref{fig:ngc_ecc} shows the energy change at the half-mass radius of
the cluster, calculated from equations (\ref{eq:dEecc},\ref{eq:adcorr}).
The value of $\langle\Delta E\rangle_h$ falls dramatically with decreasing
eccentricity, in part because $R_p$ increases as $e$ decreases and largely
because of the adiabatic corrections.  As expected, there is essentially no
heating on nearly circular orbits.  Also, we calculate the median energy
change, $\langle\Delta E\rangle_h$, for two more cases, when the orbital
energy is a half and twice the observed value.  The more bound orbits tend
to have a shorter timescale $\tau$ and correspondingly larger energy input
than the less bound orbits.

\section{Discussion}
\label{sec:discussion}

Our analysis applies to weak gravitational perturbations, tidal shocks.  A
significant evolution of a satellite or a star cluster occurs over a number
of successive shocks.  In this respect, the current work differs from that
of \cite{AW} (1985,1986) who studied the effects of strong interactions
between two similar mass galaxies.  Their work showed that the impulsive
approximation is adequate for the most regions where a significant mass loss
and energy changes occur.

\cite{AW} emphasized that particles escaping the satellite after the
perturbation should be excluded from the analysis of the remaining
structure.  In the case of tidal shocks, few stars leave the satellite
immediately.  Instead, they slowly drift in the energy space (see Figure
\ref{fig:ngc_bin}) and eventually escape through the tidal boundary.

The results of our calculations are useful for semi-analytical studies of
tidal interactions between large galaxies and their dwarf companions or star
clusters.  For example, \cite{GO:97} estimated the amount of heating of
globular clusters due to the Galactic bulge by inserting equations
(\ref{eq:int}, \ref{eq:de}, \ref{eq:masscorr}) into their Fokker-Planck
code.  Assuming the spherical symmetry of the cluster, the code solves the
Fokker-Planck equation for the distribution function of stars in energy
space.  Since the amplitude of perturbations is small, the first two terms
of the expansion of the Master equation (for example, \cite{vK:81};
\cite{G:85}) are both linear in the perturbation energy and lead to the
Fokker-Planck formulation.  The higher order terms are negligible and can be
ignored.  This approach is much faster than a full N-body simulation and
allows a direct comparison with analytical models of star clusters.  A new
detailed study of the globular cluster evolution including tidal shocks is
presented in \cite{GLO:98}.

\section{Summary}
\label{sec:summary}

We have calculated the tidal field of a spherical mass distribution.  Using
the impulse approximation, coupled with adiabatic corrections, we estimated
the amount of heating of stars in the satellite during a tidal shock.  We
considered several mass distributions for the host galaxy, a Hernquist model
and three power-law profiles, and explored various satellite orbits.
Although we restricted ourselves to spherically-symmetric galaxies to
simplify the analysis, our results should also hold for non-spherical
systems with a modest amount of flattening.

For the Hernquist model, we found that heating is enhanced by a factor of
few on eccentric orbits compared to the conventional straight path
approximation.  However, adiabatic corrections prevent any heating on
low-eccentricity orbits.  In the harmonic oscillator potential, the
satellite is equally compressed in all directions.  In the isothermal sphere
potential, heating on eccentric orbits is also larger than that in the
straight path approximation.

To illustrate the analytic results (eqs. [\ref{eq:dEecc}-\ref{eq:adcorr}]),
we considered the example of globular cluster \ngc.  For a fixed orbital
energy of the cluster, the heating is much larger on highly eccentric orbits
than on low-eccentricity orbits because of the adiabatic corrections.
N-body simulations confirm the validity of our method, which becomes a
powerful tool in calculating the tidal heating of satellites of larger
galaxies.  This semi-analytic method can also be used for studying the
evolution of giant star clusters in galactic nuclei or the disruption of
dwarf satellites in groups of galaxies.

\acknowledgements

We would like to thank Simon White and the anonymous referee for useful
comments.  This work was supported in part by the NSF under grants AST
94-24416 and ASC 93-18185, and by the Presidential Faculty Fellows Program.

\appendix
\section{Calculation of the tidal integrals for the Hernquist model}

We evaluate here the tidal integrals (eq. [\ref{eq:int}]) for the Hernquist
model.  The normalized mass profile is given by equation (\ref{eq:mbH}).
Denoting $\alpha \equiv a/R_p$, the first integral is
\begin{equation}
I_0(\alpha) = \int_1^\infty {\zeta^2 \over (\zeta+\alpha)^2} \,
  {d\zeta \over \zeta^2 (\zeta^2-1)^{1/2}}.
\end{equation}
Substituting $x = (\zeta+\alpha)^{-1}$, we have
\begin{equation}
I_0(\alpha) = \int_0^{(1+\alpha)^{-1}} {x \, dx \over [1 - 2\alpha x +
  (\alpha^2-1) x^2]^{1/2}}.
\end{equation}
Now for $\alpha > 1$, we take $y = \alpha - (\alpha^2-1)x$, and
\begin{equation}
I_0(\alpha) = (\alpha^2-1)^{-3/2} \int_1^\alpha {(\alpha-y) dy \over
  (y^2-1)^{1/2}} = -\frac{1}{\alpha^2-1} + {\alpha \over (\alpha^2-1)^{3/2}}
  \ln{(\alpha + (\alpha^2-1)^{1/2})}.
\end{equation}
For $\alpha < 1$, we have $y = \alpha + (1-\alpha^2)x$:
\begin{equation}
I_0(\alpha) = (1-\alpha^2)^{-3/2} \int_\alpha^1 {(y-\alpha) dy \over
  (1-y^2)^{1/2}} = \frac{1}{1-\alpha^2} - {\alpha \over (1-\alpha^2)^{3/2}}
  \arccos{\alpha}.
\end{equation}
It is straightforward to verify that both expressions match smoothly at
$\alpha = 1$; $I_0(1) = \onethird$. The next integral, $J_0(\alpha)$, can be
reduced to the already calculated $I_0$. Taking $x = \zeta^{-1}$, we get
\begin{eqnarray}
J_0(\alpha) & = & \int_0^1 {x^3 dx \over (1+\alpha x)^2 (1-x^2)^{1/2}} =
  \frac{1}{\alpha^2} \int_0^1 {(\alpha^2 x^2 + 2\alpha x + 1 -1 -2\alpha x)
  x dx \over (1+\alpha x)^2 (1-x^2)^{1/2}} \nonumber \\
  & = & {1-I_0 \over \alpha^2} - {2 \over \alpha} \int_0^1 {x^2 dx \over
  (1+\alpha x)^2 (1-x^2)^{1/2}} \nonumber \\
  & = & {1+3 I_0 \over \alpha^2} - {\pi \over \alpha^3} - {2I_0 \over \alpha^4}
  + {2 \over \alpha^4} \int_0^1 {(\alpha + x) dx \over
  (1+\alpha x)^2 (1-x^2)^{1/2}}.
\end{eqnarray}
Now it is useful to note that
\begin{equation}
\int_0^1 {(\alpha + x) dx \over (1+\alpha x)^2 (1-x^2)^{1/2}} \equiv 1
\end{equation}
for all values of $\alpha$.  Thus, we arrive at
\begin{equation}
J_0(\alpha) = {\alpha^2 - \pi \alpha + 2 + (3\alpha^2 - 2) I_0(\alpha)
  \over \alpha^4}.
\end{equation}
The remaining two integrals are easily calculated using the identity
(\ref{eq:deriv}), or
\begin{equation}
I_1(\alpha) = -\alpha {dI_0(\alpha) \over d\alpha}, \hspace{1cm}
J_1(\alpha) = -\alpha {dJ_0(\alpha) \over d\alpha}.
\end{equation}
For convenience, all four integrals are combined below.
{\large
\begin{mathletters}
\begin{equation}
I_0 = \left\{ \begin{array}{lr}
      \frac{1}{1-\alpha^2} - \frac{\alpha}{(1-\alpha^2)^{3/2}} \,
      \arccos{\alpha}, & [\alpha<1] \\ \\
      \frac{1}{3} \approx 0.33333, & [\alpha=1] \\ \\
      \frac{1}{1-\alpha^2} + \frac{\alpha}{(\alpha^2-1)^{3/2}} \,
      \ln(\alpha+(\alpha^2-1)^{1/2}), & [\alpha>1]
    \end{array} \right.
\end{equation}

\begin{equation}
I_1 = \left\{ \begin{array}{lr}
    -{3\alpha^2 \over \left(1-\alpha^2 \right)^2} + \frac{\alpha (1+2\alpha^2)}
    {(1-\alpha^2)^{5/2}} \, \arccos{\alpha}, & [\alpha<1] \\ \\
    \frac{4}{15} \approx 0.26667, & [\alpha=1] \\ \\
    -\frac{3\alpha^2}{(\alpha^2-1)^2} + \frac{\alpha (1+2\alpha^2)}
    {(\alpha^2-1)^{5/2}} \, \ln(\alpha+(\alpha^2-1)^{1/2}), & [\alpha>1]
    \end{array} \right.
\end{equation}

\begin{equation}
J_0 = \left\{ \begin{array}{lr}
    {-\alpha^3 +\pi \alpha^2 +2\alpha -\pi \over \alpha^3 \left(1-\alpha^2
     \right)} - {3\alpha^2 -2 \over \alpha^3 \left(1-\alpha^2 \right)^{3/2}} \,
    \arccos{\alpha}, & [\alpha<1] \\ \\
    {10 \over 3} - \pi \approx 0.19174, & [\alpha=1] \\ \\
    {-\alpha^3 +\pi \alpha^2 +2\alpha -\pi \over \alpha^3 \left(1-\alpha^2
     \right)} + {3\alpha^2 -2 \over \alpha^3 \left(\alpha^2 -1\right)^{3/2}} \,
    \ln(\alpha+(\alpha^2-1)^{1/2}), & [\alpha>1]
    \end{array} \right.
\end{equation}

\begin{equation}
J_1 = \left\{ \begin{array}{lr}
    {2\alpha^5 -3\pi \alpha^4 -11\alpha^3 +6\pi \alpha^2 + 6\alpha -3\pi
     \over \alpha^3 \left(1-\alpha^2 \right)^2} + 
     {12\alpha^4 -15\alpha^2 +6 \over \alpha^3 \left(1-\alpha^2 \right)^{5/2}}
     \, \arccos{\alpha}, & [\alpha<1] \\ \\
    9.6 - 3\pi \approx 0.17522, & [\alpha=1] \\ \\
    {2\alpha^5 -3\pi \alpha^4 -11\alpha^3 +6\pi \alpha^2 + 6\alpha -3\pi
     \over \alpha^3 \left(\alpha^2 -1 \right)^2} + 
     {12\alpha^4 -15\alpha^2 +6 \over \alpha^3 \left(\alpha^2 -1 \right)^{5/2}}
     \, \ln(\alpha+(\alpha^2-1)^{1/2}), & [\alpha>1]
    \end{array} \right.
\end{equation}
\label{eq:Hint}
\end{mathletters}
}

\section{Orbit integration}
\label{sec:A.orbits}

We describe here an accurate procedure to calculate orbits in a given
potential.  The four dimensionless potentials used in this paper and the
corresponding angular momenta are given in Table \ref{tab:pot}.  The angular
momentum of an orbit of the given perigalactic distance and eccentricity is
found by equating the orbital energy at the perigalacticon and the
apogalacticon.

We integrate the orbits using the following quadratures:
\begin{mathletters}
\begin{eqnarray}
\theta & = & \int_{r_p}^{r} \, {dr \over r^2 \left( {2\varepsilon - 2\phi(r)
  \over j^2} - {1 \over r^2}\right)^{1/2}} \\
t      & = & \int_{r_p}^{r} \, {dr \over j \left( {2\varepsilon - 2\phi(r)
  \over j^2} - {1 \over r^2}\right)^{1/2}},
\end{eqnarray}
\end{mathletters}
where $r$ is the dimensionless radius, $r \equiv R/a$ or $r \equiv
R/R_{max}$, and $\varepsilon$ is the dimensionless energy of the orbit.  To
avoid singularities at both the perigalacticon and the apogalacticon, we
separate the integrals into two parts at some intermediate radius, $r_{div}
\equiv (1+e) r_p$.  Then, we change the variables of integration
individually for the two parts to remove singularities.  For $r_p \le r \le
r_{div}$, the new variable is $u_1 \equiv \sqrt{\alpha - {1\over r}}$; for
$r_{div} \le r \le r_a$, it is $u_2 \equiv \sqrt{{1\over r} - \alpha {1-e
\over 1+e}}$.  Both parts of the integrals are regular in $u_1$ or $u_2$ and
can be calculated numerically with high accuracy.

Note, that in the harmonic oscillator potential the orbits allow simple
analytic solution.  In Cartesian coordinates,
\begin{equation}
x = r_p \, \cos{t}, \;\;\;\;\;\; y = r_a \, \sin{t},
\label{eq:A.oscorb}
\end{equation}
where
\begin{equation}
t = \arctan{\left( {1-e \over 1+e} \, \tan{\theta} \right)}.
\end{equation}

\begin{deluxetable}{lcc}
\tablehead{ \colhead{Model} & \colhead{Potential, $\phi(r)$} &
            \colhead{Angular momentum, $j^2 (\alpha,e)$} }
\startdata
Hernquist model            & $ - {1 \over 1+r}$        &
  ${(1+e)^2 \over \alpha (1+\alpha) (1+\alpha + (1-\alpha)e)}$ \nl
  \tablevspace{.2cm}
Harmonic oscillator        & ${r^2\over 2}-{3\over 2}$ &
  $\alpha^{-4} \, {(1+e)^2 \over (1-e)^2}$ \nl
  \tablevspace{.2cm}
Power-law, $n=1$           & $r-2$                     &
  $\alpha^{-3} \, {(1+e)^2 \over 1-e}$ \nl
  \tablevspace{.2cm}
Singular isothermal sphere & $\ln{r} - 1$              &
  $\alpha^{-2} \, {(1+e)^2 \over 2e} \, \ln{1+e \over 1-e}$
\enddata
\label{tab:pot}
\end{deluxetable}

\clearpage




\end{document}